IAC-22-A7.IP.2.x68449
# Concept Study for Observing Galactic Neutrinos in Neptune's Atmosphere
**Trent English [a], Dr. Nick Solomey [b]**

[a] *Department of Physics, Wichita State University, 1845 Fairmount St. Wichita, KS 67260*,
txenglish@shockers.wichita.edu
[b] *Department of Physics, Wichita State University, 1845 Fairmount St. Wichita, KS 67260*,
nick.solomey@wichita.edu

**Abstract**

I discuss the feasibility of a conceptual space-based neutrino detector that utilizes the Ice Giants as Targets for Galactic Neutrinos. The purpose of this research stems from the concept of wanting to find a new method of observing the Galactic Core (GC) of the Milky way and the Supermassive black hole, Sag A*. Observations of the GC have been made in every accessible wavelength except for the regions of space that are too dense for photons to probe. In these regions, we may instead use neutrinos. Neutrinos from the Active Galactic Nucleus are emitted at extreme energies, 10 GeV to EeV scales, but have an extremely low flux measured here at Earth. Neutrino telescopes such as the IceCube Observatory have only been able to measure a handful of neutrinos that might correlate to the GC. But using Gravitational lensing, our sun can be used as a lens which increases the "light" collection power for neutrinos by a factor of 1013, with the trade-off that the minimum focal point is located at 22 AU. This means that Uranus and Neptune are suitable natural targets for these neutrinos to interact with and observe the effects from a spacecraft in orbit. Initial studies use GEANT4, a particle physics simulation toolbox developed by CERN, to facilitate the propagation of energetic particles passing through the atmosphere of Neptune. Various aspects are studied ranging from the wavelength of the photons that are being measured at the detector, timing of the hits, and distribution of the photons leaving the atmosphere. For each of these aspects, we modify several variables such as particle type, energy, interaction depth, and orbital distance from the surface. I also discuss the versatility of this neutrino detector which has the possibility of mapping out the inner structure of the Ice Giants, in-depth studies of the neutrinos coming from the GC, and possibilities to use this method for other cosmic neutrino sources. This detector would be of great interest to planetary science, particle physics, and astrophysics communities.

**Keywords:** Gravitational Lensing, Neutrinos, High Energy Physics (HEP), Astrophysics, Galactic Core, Planetary Sciences

**Acronyms/Abbreviations**
Neutrino Solar Orbiting Laboratory: vSOL
Galactic Core: GC
Supermassive Black Hole: SMBH
Active Galactic Nucleus: AGN
Astronomical Unit: AU
Ultra-High Energy: UHE
Jet Propulsion Laboratory: JPL

## 1. Introduction
This study is a branch off of the vSOL project which is based out of Wichita State University. Now vSOL will be the first space-based neutrino detector that will be measuring neutrinos generated via fusion from the core of our sun. Various studies can be conducted such as off-axial measurements, neutrino oscillations while approaching the sun, and possible measurements for more detailed values for the upper limits to the neutrinos mass [1].

The question then arises, what if we instead go away from the sun while still wanting to look for neutrinos? The flux of solar neutrinos will decrease by a factor of $1/r^2$ as we move away from the sun and towards the outer planets. Even though our star is the brightest source of neutrinos in our sky, it is not the only prominent source that we can look at. The second brightest source is the center of the Milky Way which is located 25,000 light years from Earth [1]. Previous studies at Wichita State have investigated the concept of gravitational lensing of neutrinos from stellar objects in the galactic core [2]. Their results suggest that at the light gravitational focus of the sun, the flux of these stellar neutrinos would be 800x - 8000x the flux of solar neutrinos measured here on Earth [2]. One main note is that these calculations were only considering neutrinos that were lensed outside the surface of the sun.

This study aims to justify the feasibility of a space-based neutrino detector which utilizes the Ice Giants as possible targets for galactic neutrinos.

Various aspects have been initially studied such as particle simulations in the atmosphere of Neptune, forms of power generation of the craft, methods to increase the light collection power, and possible forms of background rejection.

### 1.1 Motivation and Background

The main motivation of this research stems from the idea of wanting to find a new way to observe and study the galactic core of the Milky Way. The GC has been a hot topic for the astrophysical community, and observations have been made in every accessible wavelength so far to better understand the chaotic environment [3]. At the center of our galaxy resides a supermassive black hole named Sagittarius A, whose characterization as an SMBH was only solidified just a few years ago [3]. Why did it take so long for us to finally prove that the center of the galaxy had an SMBH? There is difficulty that arises when trying to observe the center of the galaxy given that there is a large amount of dust, debris, and other objects between us, and the core. But there are ways to bypass the dust and debris by using telescopes in the Infrared and Radio spectrum.

Before May 12th, 2022, we have not had any clear images of what exactly Sag A* looked like given all of these obscurities. But thanks to the Event Horizon Collaboration, we now have our first detailed image of Sag A* which can be seen in Figure 1. Even though we have now produced the first detailed image of Sag A*, is there other ways that we can observe these types of regions without using photons?

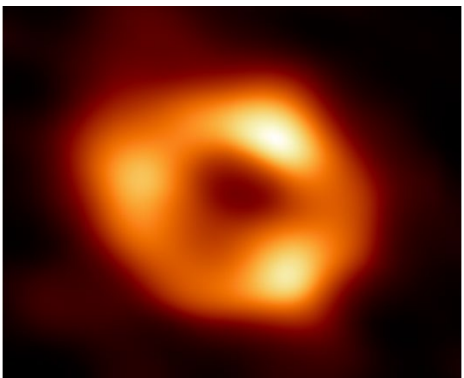

Fig. 1. First image of Sag A* produced from the Event Horizon Collaboration [4]

Aside from photons which are being produced in processes around the black hole, other particles can be ejected from this energetic environment which could be of use to us. Most of the charged particles that are produced from these environments will be deflected from magnetic fields which obscure the locations to which they were created [5]. But neutrinos that are created from an Active Galactic Nucleus, would be emitted at extreme energies and unhindered by the EM fields around them [5]. This means that we can possibly use neutrinos to help probe the surroundings of Sag A* and to have new techniques of observations.

### 1.2 Gravitational Lensing of Neutrinos

Studies have been done to look at previously collected high energy neutrino events to see if their detected tracks align with the GC. Neutrino telescopes such as IceCube have only been able to measure a handful so far that might correlate to the same region as the GC but given the spatial resolution it is difficult to say for certain [6]. As seen from Figure 2, the flux of neutrinos coming from the AGN are very low in respect to some of the other more prominent sources such as solar neutrinos. If we want to harness the neutrinos coming from the AGN, we must reduce the noise created by the other sources of neutrinos while also increasing the flux of the highly energetic contributions which can help us probe further into the GC.

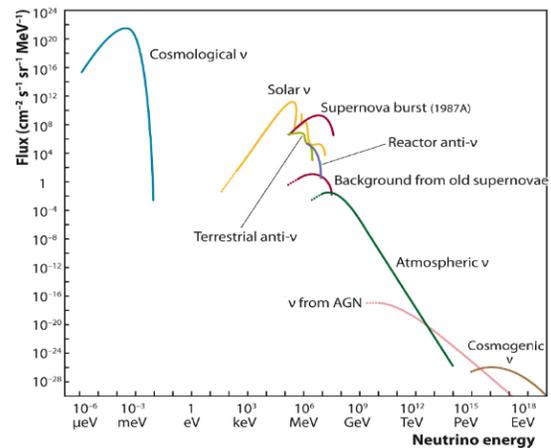

Fig. 2. Neutrino Fluxes as a function of energy and their corresponding sources [7]

To increase the flux, we can use a concept known as Gravitational Lensing which was first proposed by Einstein back in 1912 [8]. With gravitational lensing, a massive object such as star can cause light to be deflected from its original path which is a consequence of the General Theory of Relativity [9]. This means that massive objects can be used similar to that of a lens in optics which can focus light down to a ring or point depending on the observer's location relative to the focal point created by the lens. An example of this phenomena can be seen in Figure 3 which was captured by the Hubble Wide View camera in 2011.

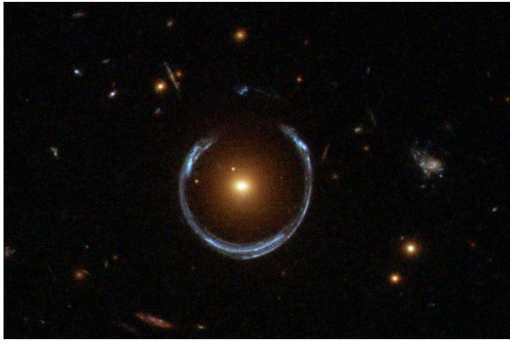

Fig. 3. The Cosmic Horseshoe which depicts the light coming from a distant galaxy being focused by a nearby galaxy [2]

Now if we can use our sun as a lens, why have we not taken advantage of this yet? For light, the gravitational focus due to our sun is an astonishing 550 AU which is the minimum focal distance of light passing the exact surface of the sun [9]. Even though the focus of our sun is located at 550 AU, the light collection power of our sun is a staggering $10^{11}$ with an angular resolution down to $10^{-11}$ arcsecs [10]. This type of magnification and angular resolution would allow us to gain valuable knowledge about celestial bodies and imaging exoplanets [10]. Now to give context to the distance we would have to go, Voyager 1 has been traveling for more than 40 years and has only travelled to 155 AU [11]. So, using our sun as a lens for light is not a task that can be done easily given the distance that we have to travel is further than any object that we have sent into space so far. But this same concept can be applied to neutrinos as well.

Neutrinos have a very small finite mass but given that it is non-zero there are differences as to how the deflection of their path occurs vs that of photons. But for the case of the particles passing outside the surface of the sun, the resulting equation that governs their deflection are the exact same. One of the main factors that help with neutrinos and gravitational lensing is that the sun is mostly transparent to neutrinos, allowing them to pass through the interior similar to that of Figure 4 [12]. Because they can pass through the sun, this means that their deflection is much greater and allows for the smallest focal point to be around 22 AU [12]. On top of the reduction in focal length, the "light" collection power for neutrinos is larger than that of photons and is on the order of $10^{13}$ which enhances our flux signal significantly [2,10].

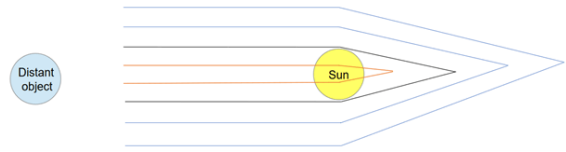

Fig. 4. Depiction of Neutrino Gravitational Lensing of the Sun.

## 2. Ice Giants as Targets for Galactic Neutrinos

### 2.1 Examples of Neutrino Detectors

Given that neutrinos interact only through the weak interaction, detection of these particles can be rather difficult. But various experiments around the world use different methods to achieve detections of these ghost particles. Depending on the type of research that is being conducted, detectors can range from large vats of water, scintillating materials, or naturally occurring media for the particles to interact with.

Super-K is a neutrino detector that uses a 50-kilotonne vat of water that detects Cherenkov radiation from neutrino interactions [13]. This detector is lined with thousands of photomultiplier tubes (PMT's) which increase the spatial resolution of where these Cherenkov bursts are being created [13]. This detector was also used to collect data of solar neutrinos and was able to construct an image of our sun using these particles. Figure 5 shows the result of their data collection, and this gives a useful and valuable concept that the same practice may be used for the GC.

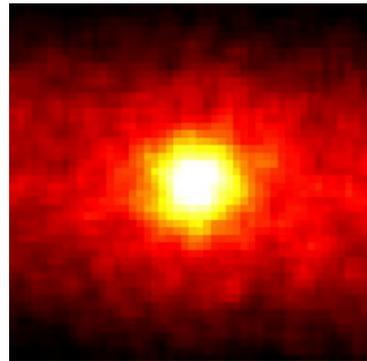

Fig. 5. Image of the sun produced by Super-K from solar neutrino events [14]

Now, a few examples of detectors that use naturally occurring media can be seen with IceCube, ANITA, and ARA. All three of these detectors are located in the South Pole and utilize the vast amounts of ice for their neutrino detection processes. The detectors for IceCube are placed into holes that are located 1.5 to 2.5 km below the surface, but the overall observatory encompasses 1 cubic km [15]. This detector has been able to measure neutrinos at energies in the PeV scale which is much higher than any particle energy that we can produce on Earth so far [15]. Just like Super-K, IceCube measures the Cherenkov radiation that is created by the charged particles but this time through the ice media.

But, both ANITA and ARA measure neutrino interactions in the ice via the Askaryan effect. The Askaryan effect is a version of Cherenkov radiation but instead of a burst of visible light that is produced it is a burst of radio waves [16]. Now this effect does not happen for low energy neutrino interactions but requires energies over 10 PeV for it to occur and must be in a dense but optically transparent media such as ice [17]. More will be discussed about the Askaryan effect later in this paper for other possible studies that can occur.

## 2.2 Benefits of Ice Giants for Galactic Neutrinos

As previously mentioned in Chapter 1.2, by using our sun as a lens for neutrinos, the minimum focal point comes down to around 22 AU. This means that Uranus is a wonderful candidate to measure the galactic neutrinos that are deflected from the sun [12]. But also given the variability of the focal lengths depending on the density approach taken and how much the neutrinos are deflected from the sun; Neptune is also a great candidate for these interactions. Now, we can use some of the information mentioned about current neutrino detection processes and the type of interaction media for this study. Given that the Earth is too close to the sun to utilize the gravitational focus, we then have to take a detector out to 22 AU.

Just like IceCube, ANITA, and ARA, we can use a natural media for these energetic neutrinos to interact with, which would be the Ice Giants themselves. Because Neptune and Uranus are primarily made of gases such as Hydrogen, Helium, and Methane they would not make for good media for the neutrinos to interact with given that they do not form a dense target. But as the energy of the neutrino increases, we also have an increased cross-section which means a higher chance for the neutrinos to interact with matter. Which this can be seen in Figure 6 that shows the resulting neutrino-nucleon cross sections for these ultra-high energy neutrinos. Meaning that even if the media is a low nucleon density, there is still a marginal chance of these interactions taking place. Experiments on Earth have made the mention that to observe some of the highly energetic neutrino processes, we would need hundreds to thousands of cubic km of water or similar mass to observe these interactions [17].

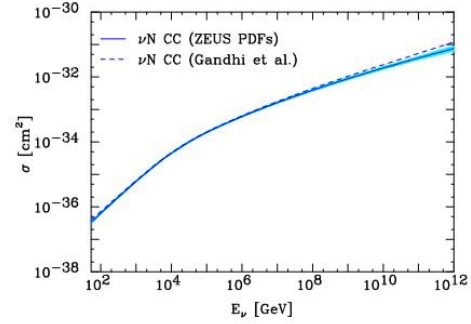

Fig. 6 Total charge current cross sections of neutrinos at UHE [18]

Now the neutrino interactions within the atmosphere of the Ice Giants can be observed given the interaction seen in Equation 1. Where l stands for the corresponding leptons, e, µ, and τ, while N would be the neutron in the nucleus and X is the converted proton [18]. Now the leptons that are created during the process would then continue through the atmosphere and given the energy ranges of the incoming neutrinos the leptons should have almost an identical trajectory. By using a spacecraft that is orbiting the planet, we can look to observe these energetic particles passing through the planet in several ways. We can observe the Cherenkov radiation, scintillation, and possibly the delayed fluorescence which would correlate to these initial galactic neutrinos.

$$\nu_l + N \longrightarrow l^- + X \qquad (1)$$

## 2.3 Alignment of the Ice Giants with the Galactic Core

Even though Neptune and Uranus both fall within the focal range of the gravitational lens of our sun, we know that these planets are not stationary and are not always aligned with the GC. This means that as the planets are orbiting the sun, they will slowly start moving into the gravitational lens and the flux of galactic neutrinos would then start to increase. Now, how often does this occur?

Uranus and Neptune will not be in the gravitational focus at the same time and are separated by several decades. Uranus will pass through the focus first around 2032 and will not spend as long in this focus due to its shorter orbit. On the other hand, Neptune will pass through the focus around 2065 and given the longer orbit would give us additional time to make measurements. Even though the Neptune alignment will be 40 years from now, it would give us a better opportunity given that we would not have enough time to build, test, and launch a craft to Uranus in that time frame. If we could send crafts out to both planets, we would then be able to look at the difference in the flux properties as the two different foci. Now if we miss these windows of opportunity, we will not get a chance again until 2116 and 2229 respectively [20,21]. But there are other cosmological sources that we can begin to study when the Ice Giants are not in alignment with the GC.

## 3. Simulations of Neptune's Atmosphere

Before we can go through and produce an image similar to that of Figure 5, we must first perform simulations to determine what type of signals we can get out of the atmosphere of Neptune. For the simulations that were performed, I used the C ++ toolbox GEANT4 (for GEometry ANd Tracking), which was developed by CERN [14]. Now GEANT4 is used to facilitate the propagation of particles through matter and the interactions that will occur [22]. This toolbox has a variety of options that can be modified to help suit the type of interactions one is wanting to study. But given how modular this toolbox can be, we must make sure that the portions of the code that we implement will be accurate for the interactions we are wanting to study.

To manipulate the results of my simulation I used the ROOT Data Analysis Framework. ROOT is another library that was developed by CERN which can work conjointly with GEANT4 to illustrate data. Results that are produced from GEANT4 can be exported as .root files which can be modified in ROOT using C code commands. We can pull portions of the data to grab information such as timing, wavelength, positions, or any other data that we have previously told the program to store into these arrays.

### 3.1 Atmosphere and Detector Construction

When it comes to modelling the atmosphere, I created concentric half sphere shells as a simple first model for the layers of Neptune. These layers were made to be 1000 km thick and have corresponding radii to match the radius of Neptune. By having the atmosphere geometry in this shell manner then there is less error when dealing with optical photons at the boundary between the two media. When creating the composition of the atmosphere, we must create our own material in the program using other predefined materials. The composition of the outer atmosphere of Neptune is 80% Molecular Hydrogen, 19% Helium, and around 1% Methane by volume [20]. But other characteristics can be defined to increase the accuracy of our results such as current state of matter of the material, pressure, and temperature, which alter the energy deposition of particles through the material.

The material of our detector is of no importance for our detections in this simulation because we count the number of photons that pass into its volume. But the geometry was made into an arc that wraps around the shells and collects the light leaving the atmosphere.

We can then mimic what a detector of a specific size would measure from all the previously collected data.

### 3.2 Approximations of the Atmosphere

Now, one of the more notable peculiar features is that with GEANT4 if you are wanting to simulate optical processes you must define various optical properties for them to work accordingly [22]. This means if you don't know information about the index of refraction, attenuation length, and scintillation properties then the program will not produce optical photons [22]. But given that we have not been back to the Ice Giants since the Voyager missions, we have not been able to do these additional tests to acquire those properties meaning that we must make certain assumptions. The first property that we can go through and approximate is that since the atmosphere is 80% Molecular Hydrogen, then we can use the index of refraction of hydrogen gas for our atmospheric material that we have defined. We may use Equation 2 to find the index of refraction at various wavelengths, where n is the index of refraction, and $\lambda$ is the wavelength in units of micrometres [24]. Using this equation, we can determine the index of our material at various wavelengths and program that into GEANT4.

$$n - 1 = \frac{0.0148956}{180.7 - \lambda^{-2}} + \frac{0.0049037}{92 - \lambda^{-2}} \qquad (2)$$

The next characteristic that we had to assume is the attenuation length of the atmosphere. Certain groups have performed calculations to determine the mean opacities of giant planets and ultracool dwarfs [25]. From the opacities that they had calculated, I was able to determine a reasonable value for the attenuation length of the atmosphere. Now the tables that they produced were separated out depending on the metallicity of the star for that region, temperatures, densities, and pressures. Based on the pressure and density of Neptune's atmosphere, I was able to find a value for the Rosseland mean opacity, $\kappa_R$, of $9.989 * 10^{-3}$ $cm^2/g$ [20,25]. This opacity was then converted into an attenuation length of roughly 2200 meters.

Another important thing to note is that GEANT4 does have neutrinos built into the simulation, but they do not have any form of interaction capabilities. They are created in the form of by-products of reactions to carry away the excess energy and momentum. I will assume that the leptons that are produced carry away all of the energy from the interaction. This means that I only have to worry about firing each of the possible leptons, $e^-$, $\mu^-$, and $\tau^-$, through the atmosphere and observe the effects of their propagation.

## 4. Results of Simulations

The simulations were run over several loops for different particle types, energies, interaction depths, but also orbital distance. By looping over each of the parameters, this helps us to characterize and understand what type of signals we can see from the galactic neutrino. Given the number of parameters and subtopics that I studied, not all plots generated will be put into this manuscript. Additional plots can be seen in Appendix A that show Muon and Tauon data. But a link to a GitHub repository has additional plots of the various orbits and energies that can't be presented in this paper.

### 4.1 Cherenkov Hit Distributions

One of the main processes that we can observe from the atmosphere of the planet will be in the form of Cherenkov radiation. The plots in this section show the results of how the Cherenkov distributions vary by orbital distance, and examples of other particle types can be seen in Appendix A. But then, we see how this distribution appears from a spacecraft that has an diameter of 10 meters and is transversing over the direct path of the neutrino leaving the atmosphere. Lastly, I wanted to look at how long these processes can last in the atmosphere while the particles are passing through.

### 4.1.1 Total Distribution Hit Map

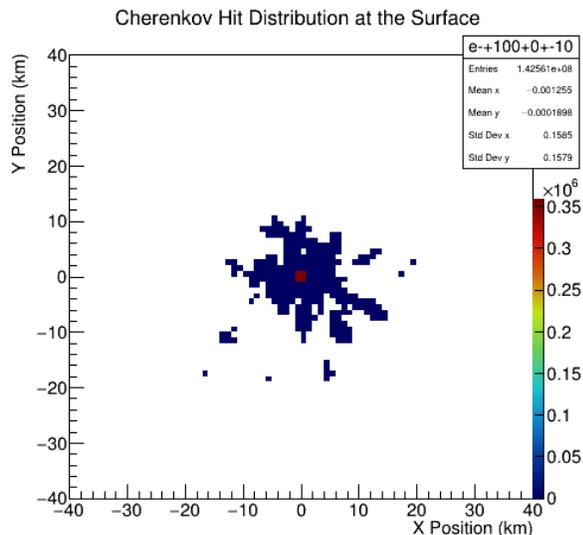

Fig. 7 Cherenkov hit distribution generated by a 100 GeV Electron at the surface of the atmosphere.

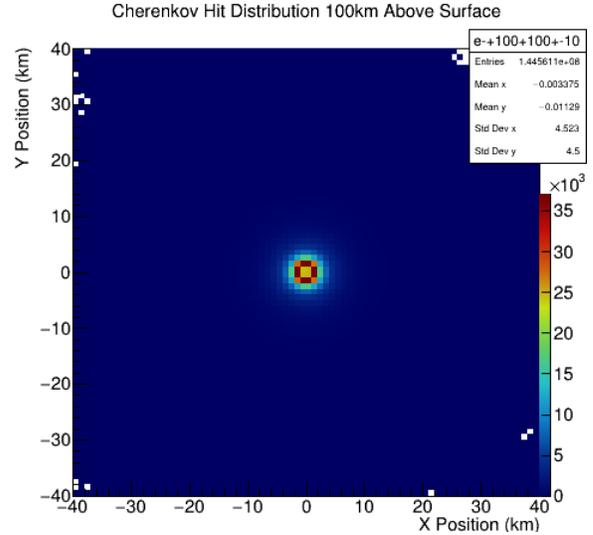

Fig. 8 Cherenkov hit distribution generated by a 100 GeV Electron at 100km from the atmosphere.

### 4.1.2 Orbital Transit Distributions

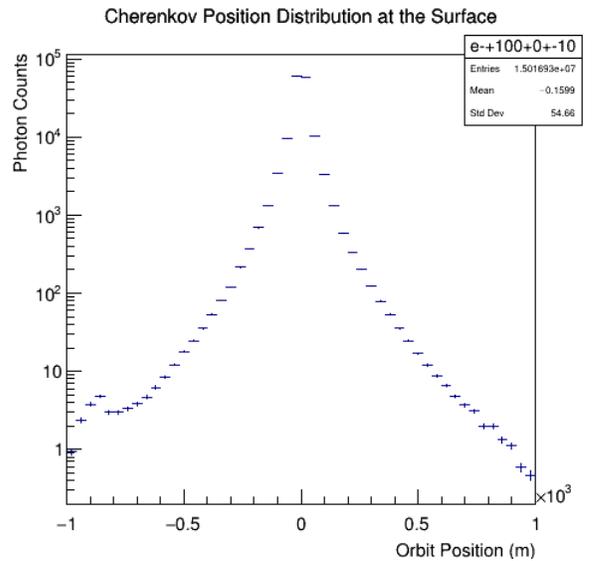

Fig. 9 Cherenkov distribution generated by a 100 GeV electron as seen from a 10-meter-wide detector traversing the surface of the planet.

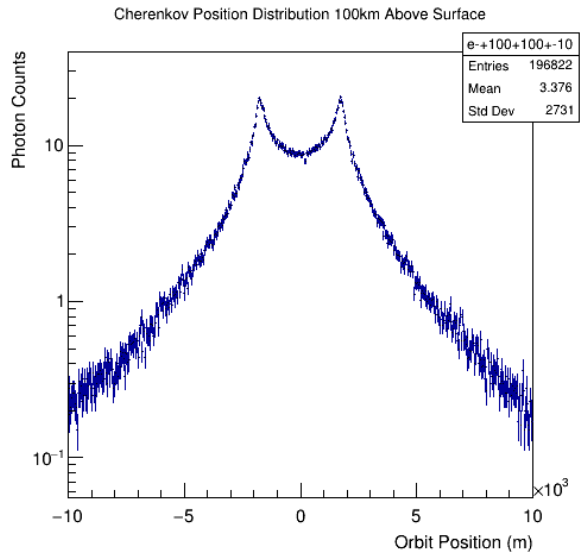

Fig. 10 Cherenkov distribution generated by a 100 GeV electron as seen from a 10-meter-wide detector traversing 100km above the surface of the planet.

*4.1.3 Timing of Detections*

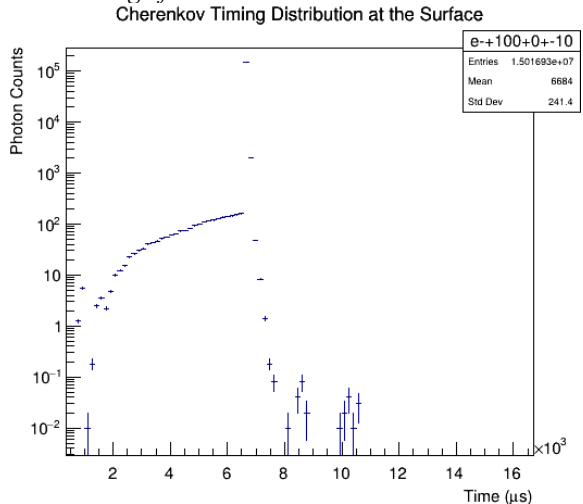

Fig. 11 Timing distribution of the Cherenkov light leaving the atmosphere as seen from a 10-meter-wide traversing the surface of the planet.

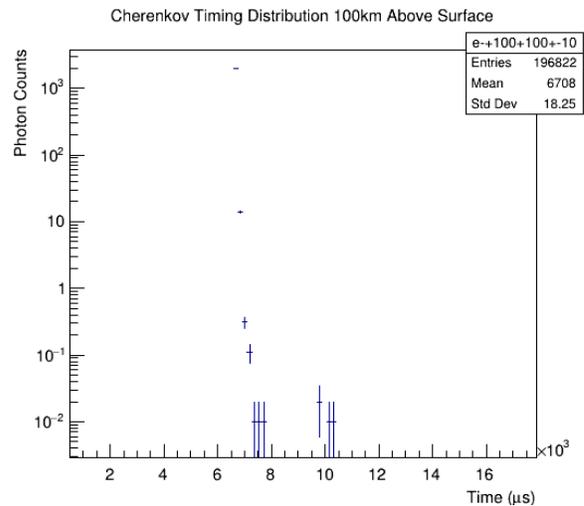

Fig. 12 Timing distribution of the Cherenkov light leaving the atmosphere as seen from a 10-meter-wide traversing 100km above the surface of the planet.

*4.2 Previous Scintillation Measurements*

During initial simulation studies, the scintillation properties of Neptune were modelled after the fluorescence properties of Earth's atmosphere. Earth has a fluorescence yield of about 5 Photons per MeV, meaning that if we have a 10 GeV particle deposit all its energy into the atmosphere, we only get 50,000 photons that are created. The results of these simulations demonstrated that the scintillation yield of the atmosphere would not create a measurable signal for our detection process. But since little information is known about the optical properties of Neptune, more studies must be conducted to understand this environment.

## 5. Discussions
### 5.1 Detector Design

Another study of this research is looking into various parameters of the spacecraft, such as power generation, overall size of the detector, but also concepts governing the detection process. As previously mentioned, the alignment of the Ice Giants with the Sun and GC is relatively short with a duration only lasting 1-2 years. If we piggyback onto other large-scale missions that are already planning to go out to the Ice Giants in these time frames, then we can perform our measurements but at the cost of the size of our spacecraft. I will discuss in this chapter that even with a small-scale spacecraft, such as a CubeSat, we can still perform the necessary measurements of these neutrino events.

*5.1.1 Concept of Detection Process*

Given the variety of different sources of light that can be produced between the neutrino events and natural occurrences in the atmosphere, we need a detector concept similar to that of Figure 13. Now this detector would have multiple CCD cameras and sensors to look at all possible wavelengths that are coming out of the atmosphere and that are incident on our device. Lightning produced in the atmosphere would emit a large number of photons across all wavelengths but have very prominent signals in the infrared spectrum. Thus, meaning if the detector were to see signals in all of the sensors but also a large signal in the IR sensor then we would reject that event because it did not come from a neutrino interaction. Similar ideas can be applied to Auroras since they can produce a wide array of wavelengths, but we would have to study and characterize precisely the type of signals that they would emit.

For the Cherenkov radiation that is created during these neutrino interactions, they produce large amounts of light in the UV and blue portion of the spectrum. Filters can be applied to each of the sensors to pick out regimes of light to help reject events. Another process which was not looked at in the simulations but can also be a contributing factor could be delayed Fluorescence that is caused by these neutrino interactions. The elements in the atmosphere could be put into an excited state and at some time later will return to the ground state but in that process will emit energy in the form of light. This can be used in tandem with the other processes to be identifiers to help prove that the signals that we saw were from these galactic neutrinos. The combination of selection criteria can be observed in Table 1 which outlines what to look for in each process.

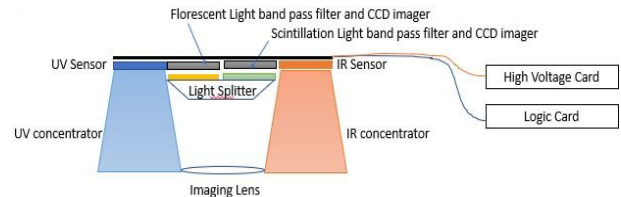

Fig. 13 Schematic of detection process to distinguish between neutrino events and background signals.

Table 1. Identification markers for possible detection and rejection of signals

| Process Type | Spectrum | Timing | Direction | Selection Criteria |
|---|---|---|---|---|
| Cherenkov | Blue | Prompt | Forward Cone | Initiate with Blue Signal |
| Scintillation | TBD | Fast | Isotropic | Delayed Signal to Observe |
| Fluorescence | TBD | Slow | Isotropic | Delayed Signal to Observe |
| Lightning | All Wavelengths | Prompt | Isotropic | Veto with Large IR |
| Aurora | TBD | Continuous | Isotropic | Veto Events with continuous hits |

*5.1.2 Light Collection Power*

Previous simulations that were conducted showed that a detector that is between 5 to 10 meters in diameter would acquire appreciable detections from these interactions. But even telescopes like Hubble only have primary mirrors that are only 2.4 meters wide. I also mentioned that this type of craft would be something the size of a CubeSat, so how does any of this information match together? Well, our primary craft would have a structure that is similar to a 3U CubeSat meaning that it would be about 30cm x 10cm x 10cm. This would be the housing for our primary detector system that I previously discussed, battery packs, electronics, and any other science payload that we wish to have. To achieve the light collection power of a 5-to-10-meter size detector, we can use a concept that is being pursued by JPL.

JPL's idea is to encapsulate their satellites within an inflatable bubble to create an artificial environment, but half of the bubble will be coated with a reflective layer [26]. The reflective layer has a multipurpose functionality to it, given the larger area that it creates and being able to reflect light, it can enhance the intensity of the sunlight when the spacecraft is further from the sun. This allows some satellites or other spacecrafts to still use solar as a form of power generation for their electrical systems for some deep space missions [26]. Also, it can be used for communications systems since it would act as a large dish which enables the spacecraft to easily capture and transmit radio signals to and from Earth. In both instances, this system helps to reduce the overall mass of the craft given that it does not have to take either a massive form of power generation or a larger conventional radio transceiver.

For our case, we can use the same concept to increase our light collection power and allows us to gather a significant portion of the light leaving the atmosphere. A diagram of this concept can be seen in Figure 14, where the CubeSat would reside at the center of this bubble environment, and as light rays enter the bubble, they will hit the reflective layer and be redirected to our detector. This bubble system then allows us to have a very small spacecraft with immense light collection power, meaning we do not need a telescope more massive than Hubble in order to perform these experiments.

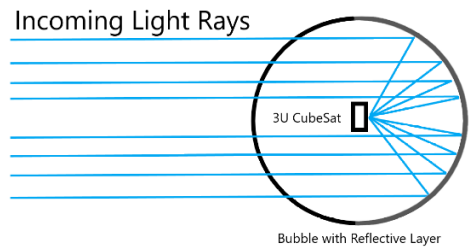

Fig. 14 Diagram of reflective bubble environment redirecting the incoming light rays back to the CubeSat.

### 5.1.3 Power Generation

Given that this will be a smaller CubeSat that is running things such as communications, flight computer, GPS system, and CCD cameras with other optical devices, we will not be requiring a large amount of power consumption. But using the standard form of power generation, i.e solar panels, would not be ideal in our case when we are orbiting Uranus or Neptune. This is because the solar irradiance measured out at Neptune would 900 times smaller than what we measure here at Earth.

Another reliable source of power generation can be from an electrodynamic tether (EDT). An EDT utilizes the magnetic field of the planet and the process of induction to power a spacecraft [27]. As our spacecraft orbits the planet, it will be traversing through the magnetic field lines and with a conductive tether attached to our craft, a current will be induced in the tether which can charge our batteries. Studies have shown that with a 10-kilometer tether attached to a craft that is orbiting Earth can produce on average 1kW of power [27]. Now this tells us that for every meter a tether can produce 0.1W of continuous power, but what about when we are orbiting Neptune? The magnetic field of Neptune is 27 times that of Earth, meaning that the previous ratio I mentioned would then be 2.7 W/m [28]. This type of power generation on its own would be sufficient to power our CubeSat, especially if the majority of our components are powered down when we are on the sun facing side of the planet.

One benefit, which can be seen as a disadvantage of this EDT system is that the induction will cause an opposing force on our craft which would slowly pull us into the planet. But orbital studies would need to be conducted to determine exactly how long it would take the orbit to degrade.

### 5.2 Versatility

The primary motivation of this research was to find new ways that we can observe the GC of the Milky Way and probe Sag A* in more detail than before. First, I had to determine if it was possible to observe these neutrino events leaving the atmosphere of Neptune, and what type of distributions do they produce. Now this type of detector can be used in more ways than just to image the GC, but still requires these highly energetic neutrinos and gravitational lensing to do so. In this chapter, I discuss the versatility of this spacecraft which can possibly be used to map out the inner structure of the Ice Giants, image various Pulsars that align with our sun, but also to conduct studies of neutrino oscillations and refinements of the upper limit to the mass of the neutrinos.

### 5.2.1 Mapping the interior of the Ice Giants

One of the main areas of science that could benefit from this type of detector would be in the realm of planetary sciences. There is great interest that comes with studying the outer planets, especially when we are wanting to determine how the structure of the core compares to our current predictions. Now a current model of the structure of Neptune can be seen in Figure 15, which illustrates that the core should be a mixture of rock and ice, while the mantle is more of water/ice media. Gathering images of the inside of the planet is difficult given the large distance that light would have to travel through the planet. But also, sending probes into Neptune would present another challenge given that the winds of Neptune can reach upwards of 2000 km/hour [28]. Thus, determining a new method of observation would provide a valuable resource that can be applied to both of the outer planets.

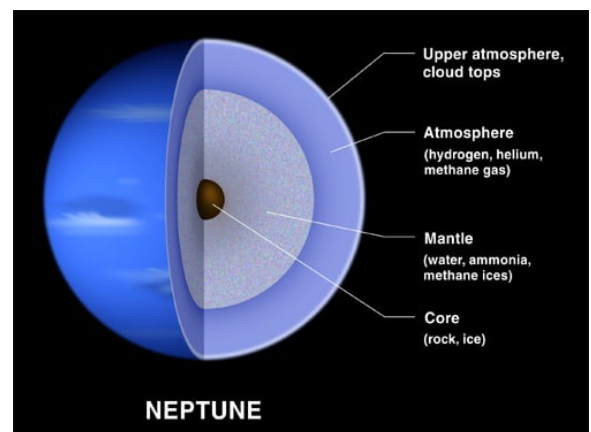

Fig. 15 Sketch of the Internal Structure of Neptune produced by NASA

An alternative form of studying the inner core of the Ice Giants can be from these energetic neutrinos that come from the GC or other prominent neutrino sources. Now the electron simulations showed that we would only get noticeable signals if they were produced in the last 50 km of the atmosphere given the amount of energy they would deposit as they transverse through the media. Meaning that the electron neutrinos interactions would not be able to measure the inner structure of the planets. But the muon and tauon particles that were sent through and tested have measurable results while being produced at much greater depths than that of the electrons. Given the longer lifetime and lower rest mass of the muon, we can use this particle to possibly probe deeper into the planets core while still being able to measure the signals from our craft orbiting the planet.

Now another very interesting concept that can possibly be used to map out the inner structure of the core comes from the Askaryan effect. As previously mentioned in Chapter 2.1, the Askaryan effect is a form of Cherenkov radiation but produces radio waves instead of visible light. Now as these ultra-high energy neutrinos interact with the ice media in the core of the planet, a cascade effect can occur that would allow for a burst of radio frequencies to be emitted from the interaction. Just like the Cherenkov radiation that was studied for this research in the visible light spectrum, we could also have these rings of radio waves that we could look for at the same time. One of the main draw backs to this process is that this only occurs for energies above 10 PeV. Given that these UHE neutrinos are very unlikely to pass through the inner core of the sun, it would be extremely rare to measure these type of radio bursts. But if refinements to neutrino studies demonstrate that there is a probability of them being measured in the range of Uranus or Neptune, then this process becomes a viable option that we can use.

### 5.2.2 Imaging Pulsars

Now the GC and the sun are two of the most prominent neutrino sources in our sky, but they are not the only sources that we can attempt to observe. Determining the origin of these neutrino sources can be a challenge as mentioned from the IceCube experiment which tried to find a correlation between their detected neutrino tracks and the GC. But other ideas have also investigated the correlation between gamma ray bursts and neutrino events as well. We can use images like Figure 16 to find various gamma ray sources and to begin a search for other neutrino sources that can be used with gravitational lensing.

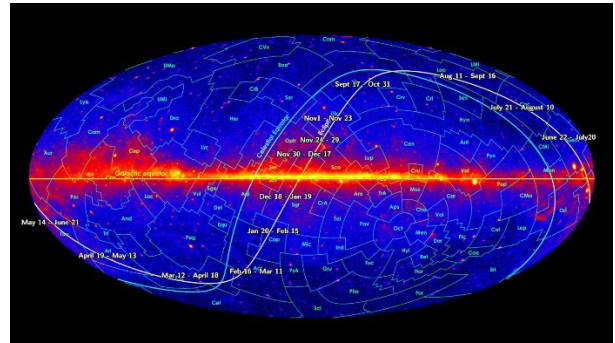

Fig. 16 Figure produced from the Fermi Gamma Ray Telescope with overlaying Celestial Equator, Ecliptic, and corresponding dates of Earth's transit [29]

When looking over Figure 16, the red portions of the picture correspond to gamma ray measurements that have been made over a five-year span. We can see that a large majority coincide to the galactic equator where the UHE neutrinos that I previously studied are being emitted from. But there are other portions that are off the galactic equator and are still very prominent in their signal strengths. Now the Ecliptic line that passes through the image has corresponding dates of where Earths position is located during its orbit. The Earth passes the galactic center near the beginning of December, but if we look back towards the May-June time frame we see other sources that fall almost perfectly on the Ecliptic line. These two bright sources from the top to the bottom are the Geminga and Crab Pulsars respectively. Studies have suggested that while Pulsars emit these highly energetic gamma ray bursts, they would also emit muon neutrinos in the TeV energy range given the energetic charged pions that this environment can produce [30].

The rate of measuring neutrinos coming from the Crab Pulsar on Earth are on the order of 0.009 km$^{-2}$yr$^{-1}$ [30]. This rate is much lower than the solar and galactic neutrinos by orders of magnitude but using the gravitational lens of our sun we would be able to increase the flux to something more reasonable to measure. We saw that for Earth's transit that it would align with these Pulsars about half an orbit after the alignment with the GC. This means that Neptune and Uranus would follow the same concept suggesting that they should be aligned with these Pulsars around 2147 and 2074 respectively. Now if researchers miss the chance to measure the galactic neutrinos that they still have these other sources which can be observed using the gravitational focus of our sun.

### 5.2.3 Neutrino Measurements

This type of detector could also be used to perform measurements of the characteristics of these neutrinos as well. One of the main studies that could be performed would be determining what fraction of the neutrino interactions occurred from the electron, muon, or tau neutrinos. These types of measurements would provide data needed for neutrino oscillation experiments that would have a theoretical framework of the type of neutrinos being produced from the energetic environments we are aligned with. Then given that we have very precise measurements of the distances from these sources, we would have an idea of what fraction of neutrino flavours that we should measure in the atmosphere. This again would tie into the concept of data that can be useful for refining the upper limit of the mass of the neutrino.

### 5.3 Future Work and Refinements

To further advance the validity of this project, more funding must be acquired for separate independent studies. Understanding the optical properties of Neptune's atmosphere is of great importance given that most of the approximations in the simulations are from these unknowns. A gas mixture can be acquired following the same chemical composition of the outer atmosphere, and studies can be conducted to determine the scintillation properties, index of refraction, and attenuation length. From these updated optical properties, we can refine the simulations to give more accurate results of what we should observe from the interactions. But in the simulations, we can go through and apply the detection process that was mentioned in chapter 5.1.1 and including background signals to demonstrate the rejection capabilities.

### 6. Conclusion

Now this research is the first steppingstone into a possible mission out to the Ice Giants to conduct these neutrino experiments. But before we can do that we must continue with our preliminary studies and simulations to better refine the results that our spacecraft would measure. This type of detector would open up new opportunities for neutrino experiments given the limitations that all current detectors are fixed at Earth. While vSOL will be the first space-based neutrino detector, it will only be focused on the solar neutrinos as it flies towards the sun. The detector concept laid out in this paper would be the first space-based neutrino detector wanting to go away from the sun and focus on these galactic neutrinos and other prominent sources.

### 6.1 Simulation Studies

The results seen in this paper and from the additional plots demonstrate that electron interactions through the atmosphere do produce significant amount of Cherenkov radiation and are well behaved along the initial trajectory. Muon and Tauon distributions do present interesting behaviours given their resulting decays, low amount of Cherenkov radiation emitted, and can deviate from the initial trajectory. Now the muons do have the benefit that they can be produced much deeper than that of the electron, but their resulting decay into an electron allows for these bursts of photons near the surface which can be measured.

### 6.2 Detector Concepts

The concepts that were laid out for our detector design present valuable information that allows for this type of craft to be possible. By creating a rejection process scheme, the detector can filter out background signals that could be caused by natural events in the atmosphere but also making sure that we can characterize these neutrino events. Given that the light collection size is an astonishing 5 to 10 meters it seemed that the spacecraft would need to be a massive structure in order to perform the measurements. But using the bubble environment that is being studied by JPL, this brings the detector to a feasible mass and within our reach to perform a demonstrator mission. But also, by utilizing the magnetic field of Neptune as a continuous source of power, we don't have to rely on solar or radioactive sources that would add additional mass.

### 6.3 Versatility

As mentioned previously, the versatility of this craft opens up wonders for different regions of science such as planetary, astrophysics, and particle physics. Given that this concept could be applied to imaging any prominent neutrino source, we have the possibility of creating images similar to that of Figure 5 for various celestial objects. Also, by having an increase of neutrino events from these stellar sources we can acquire more data relating to neutrino oscillation measurements and refinements to upper limit to the neutrinos mass. One of the main takeaways that can be used with this detector is the possibility of mapping out the inner structure of the Ice Giants. As I mentioned before, the muons can be produced at greater depths of that of the electrons meaning that they can be used to probe the inner structure of the planets. In tandem to the muons, if neutrinos of 10 PeV and above interact with the core we can utilize the Askaryan effect and measure the radio bursts emanating from the center of the planet.

These measurements would give some of the first direct insights into the structure of the Ice Giants.

**Appendix A (Results Figures)**
The link to the GitHub repository is listed below that will allow you to view all of the data collected so far from these simulations.

https://github.com/NeuGravIGMission/NeuGravMissionPlots_IAC.git

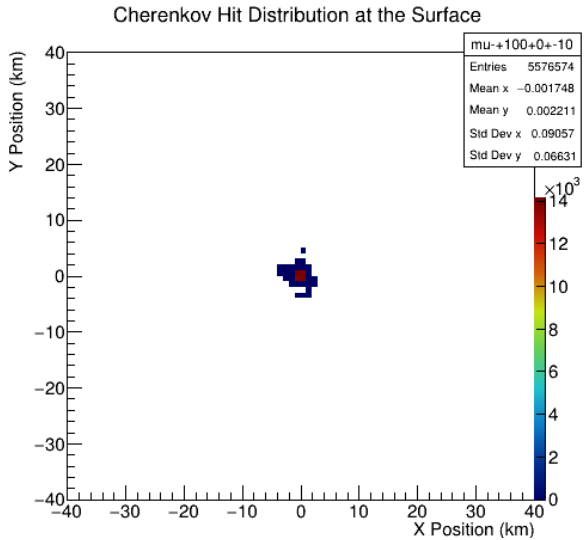

Fig 17 Cherenkov hit distribution generated by a 100 GeV Muon at the surface of the atmosphere.

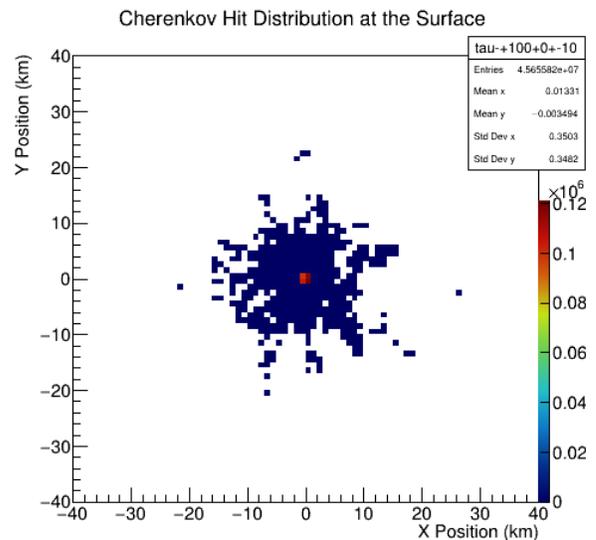

Fig 19 Cherenkov hit distribution generated by a 100 GeV Tauon at the surface of the atmosphere.

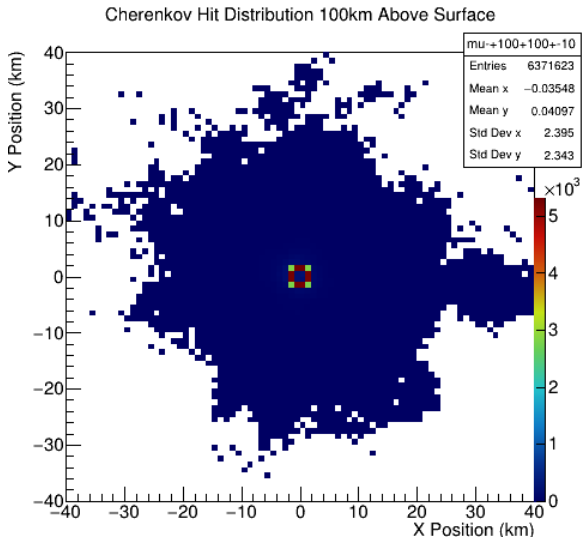

Fig 18 Cherenkov hit distribution generated by a 100 GeV Muon at 100km from the atmosphere.

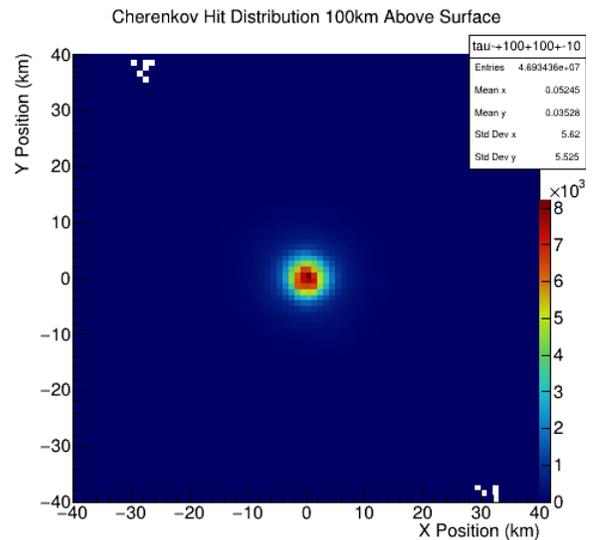

Fig 20 Cherenkov hit distribution generated by a 100 GeV Muon at 100km from the atmosphere.